\documentclass[twocolumn,showpacs,preprintnumbers,amsmath,amssymb]{revtex4}
\usepackage{amsmath}
\usepackage{graphicx}
\usepackage{dcolumn}
\usepackage{amssymb}
\usepackage{amsfonts}
\usepackage{latexsym,epsfig}

\makeatletter
\newenvironment{tablehere}
  {\def\@captype{table}}
  {}

\newenvironment{figurehere}
  {\def\@captype{figure}}
  {}
\makeatother

\renewcommand{\log}{\,\mbox{ln}}
\def\beq{\begin{eqnarray}}
\def\eeq{\end{eqnarray}}

\begin{document}

\title{Discriminating among the theoretical origins of new\\
heavy Majorana neutrinos at the CERN LHC}
\vskip 6mm
\author{F.~M.~L.~de Almeida Jr., Y.~A.~Coutinho, \\
J.~A.~Martins Sim\~oes, A.~J.~Ramalho, S.~Wulck}
\affiliation{Instituto de F\'isica, \\
Universidade Federal do Rio de Janeiro,\\
Rio de Janeiro, RJ, Brazil}
\email{marroqui@if.ufrj.br, yara@if.ufrj.br, simoes@if.ufrj.br, ramalho@if.ufrj.br, steniow@if.ufrj.br}

\author{M.~A.~B.~do Vale}
\affiliation{Departamento de Ci\^encias Naturais,  
Universidade Federal de S\~ao Jo\~ao del Rei, \\
S\~ao Jo\~ao del Rei, MG, Brazil }
\email{aline@ufsj.edu.br}
\vskip 15mm

\date{\today}
\begin{abstract}
\par
A study on the possibility of distinguishing new heavy Majorana neutrino models at LHC energies is presented. The experimental confirmation of standard neutrinos with non-zero mass and the theoretical possibility of lepton number violation find a natural explanation when new heavy Majorana neutrinos exist. These new neutrinos appear in models with new right-handed singlets, in new doublets of some grand unified theories and left-right symmetrical models. It is expected that signals of new particles can be found at the CERN high-energy hadron collider (LHC). We present signatures and distributions that can indicate the theoretical origin of these new particles.  The single and pair production of heavy Majorana neutrinos are calculated and the model dependence is discussed. Same-sign dileptons in the final state provide a clear signal for the Majorana nature of heavy neutrinos, since there is lepton number violation. Mass bounds on heavy Majorana neutrinos allowing model discrimination are estimated for three different LHC luminosities.
\end{abstract}

\maketitle


\section{Introduction}
\par
{There are a lot of very convincing experimental data for neutrino oscillations and non-zero masses, \cite{PDG}. The smallness of neutrino masses \cite{PDG} is generally understood as a consequence of some see-saw mechanism. This brings the question of the possibility of new heavy neutrino states. So far none of these new states was experimentally observed \cite{PDG,ZUB} with masses up to $M_N\simeq 100$ GeV. For higher masses there are many suggestions of experimental possibilities in the next high energy hadron-hadron colliders \cite{YSP,MAJ}, in electron-positron linear accelerators \cite{ARS,DJO,ACV} and in neutrinoless double-beta decay \cite{BIL}.  The properties of these new heavy states are the central point in many of the theoretical models proposed as extensions of the present Standard Model of elementary particle physics. Heavy Majorana neutrinos naturally generate B-L asymmetries and are good candidates for leptogenesis \cite{NAR,BUC}.  Besides the masses and mixing angle values, an important point is the Majorana or Dirac nature of new heavy neutral states. This is directly connected with lepton number conservation and the general symmetries of any extended model. 
\par
The authors already investigated some points in this direction  for a single heavy Majorana
production in hadron-hadron colliders \cite{MAJ}. It was shown that this mechanism could be
more important than pair production and that like-sign dileptons can give a clear 
signature for this process.
In the present paper we consider more general possibilities for a  heavy neutral
lepton production at hadronic colliders. If a new heavy neutral lepton is experimentally 
found, then a fundamental question to be answered will be its theoretical origin - singlet (VSM)
 \cite{VAL}, doublet (VDM) \cite{VDM}, mirror (FMFM) \cite{FMF} and other extensions.
\par
The main point of our work is to present a study of signatures and 
distributions that could, in principle, show the fundamental properties
of possible new heavy neutral leptons at the LHC energies. During the first years, the LHC Collaborations will concentrate theirs efforts on new phenomena analyzing channels with low multiplicity and the simplest SM extensions. 
Since we presently have no signal for new additional gauge bosons, we will make the hypothesis that the dominant interactions of the new heavy neutrino states are given by the standard $SU_L(2)\otimes U_Y(1)$ gauge bosons. We can resume the new particle interactions in the neutral and charged current Lagrangians:

\begin{equation}
{\cal L}_{NC}=-\sum_{i}\frac{g}{2c_W }\overline{\psi_i}\gamma^{\mu}
\left(g_V-g_A\gamma_{5}\right)\psi_i Z_{\mu}.
\end{equation}

and

\begin{equation}
{\cal L}_{CC}=-\sum_{i,j}\frac{g}{2\sqrt{2}}
\overline{\psi_i}\gamma^{\mu}\left(g_V^{ij}-g_A^{ij}\gamma_5\right)\psi_jW_{\mu},
\end{equation}
where $i,j$ stand for the usual fermions and the heavy Majorana neutrino ($N$). These models have only two parameters: the new Majorana neutrino mass and a mixing angle, if we consider all the mixing angles with the same value.
\par

The general fermion mixing terms are given in \cite{ARS} as well as the experimental bounds on the mixing angles ($\sin \theta_{mix} \equiv s_m $). For the purpose of the present investigation we resume these mixing angles in Table 1. The $N-e-W$ vertex  is of pure $V-A$ type for the VSM, pure $V+A$ type for the VDM and pure vector $V$ or pure axial $A$ for the FMFM.  
\par

The study of the bounds on the mixing angles can be addressed in several ways. A general consequence of fermion mixing is that the standard model couplings of gauge bosons and fermions is to be multiplied by a factor of the form  $\cos \theta_{mix}$. The universality of the different families coupling to the gauge bosons then imply a global upper bound on mixing angles that must all be very small, regardless of the studied model.  The present value, with $95\%$ C.L., for the upper bound on the mixing angle $\sin^2 \theta_{mix}$, is $0.0052$, when considering them all equal. A detailed analysis of these calculations can be found in \cite{MAJ}.
Another kind of analysis on mixing angles can be done for the experimental discovery potential at the LHC. In this case, besides the mixing angles one must add to the analysis at least one new unknown parameter - the heavy neutrino mass. This kind of approach was recently presented in reference \cite{HAN} for the $V-A$ case. The possibility of new gauge interactions can add more unknown parameters to the analysis. The other $ V+A, V, A $ cases can be studied in a similar form. In this paper all the mixing angles have been considered to be equal to the upper bound of the global approach.

This study is organized as follows: in section II we present the total cross-sections for pair and single production for different models of heavy Majorana neutrinos at the proton-proton collisions at LHC energies, and the various distributions that were studied trying to distinguish models, as well as the characteristics and number of events of the chosen processes. In section III, we discuss the statistical treatment used to disentangle models and determine the theoretical origin of a heavy Majorana neutrino, if experimentally found. Finally, we present our conclusions.

\par
\begin{table*}\label{eins}
\begin{tabular}{||c|l|l|l|l||}  \hline\hline
  & VSM  & VDM  & FMFM (Axial) & FMFM (Vector) \\  \hline
$W\longrightarrow e N$ & $g_V=s_m$ &  $g_V=-s_m$ & $g_V=0$  & $g_V=2s_m$\\
 & $g_A=s_m$ & $g_A=s_m$ & $g_A=2s_m$ & $g_A=0$\\ \hline
$Z\longrightarrow \nu N$ & $g_V=s_m$ & $g_V=-s_m$ & $g_V=0$ & $g_V=2s_m$\\
& $g_A=s_m$ & $g_A=s_m$ & $g_A=2s_m$ & $g_A=0$\\ \hline
$Z\longrightarrow N N$ & $g_V=s^2_m$ & $g_V=2$ &  $g_V=1$ &  $g_V=1$ \\
 & $g_A=s^2_m$ & $g_A=s_m$ & $g_A=-1$ & $g_A=-1$\\ \hline\hline

\end{tabular}
\caption{Mixing angles for $W$ and $Z$ couplings with new neutral leptons for VSM, VDM and FMFM.}
\end{table*}

\section{Total Cross-sections and Distributions}
\par

In our tree level calculations, we have implemented the models in the program CompHEP \cite{PUK} and have used the following cuts to take into account the detector acceptance: $\vert \eta_e
\vert \le 2.5$ (pseudo-rapidities of the final electrons); $\vert \eta_W \vert \le 5$ (pseudo-rapidity of the $W$); $E_e > 5$ GeV (electron energies), and $M_{ee} > 5$ GeV (invariant masses of the pairs of final electrons).

The total cross-section for single and pair production of heavy Majorana neutrinos at the LHC energy is given in Figure 1. As a consequence of the mixing terms, for the $V-A$ case the pair production is strongly suppressed. Pair production is the dominant mechanism for pure vector and axial couplings. We found no sensible difference between axial FMFM and vector FMFM for the total cross sections and for the distributions studied. So hereafter we will mention FMFM for both cases. For the $V\pm A$  case, the dominant process is the single heavy neutrino production. Using a luminosity of $100$ fb$^{-1}$, the upper bound on the Majorana mass is $500$ GeV for the single production case. 

\begin{figure} 
\includegraphics[width=0.5\textwidth]{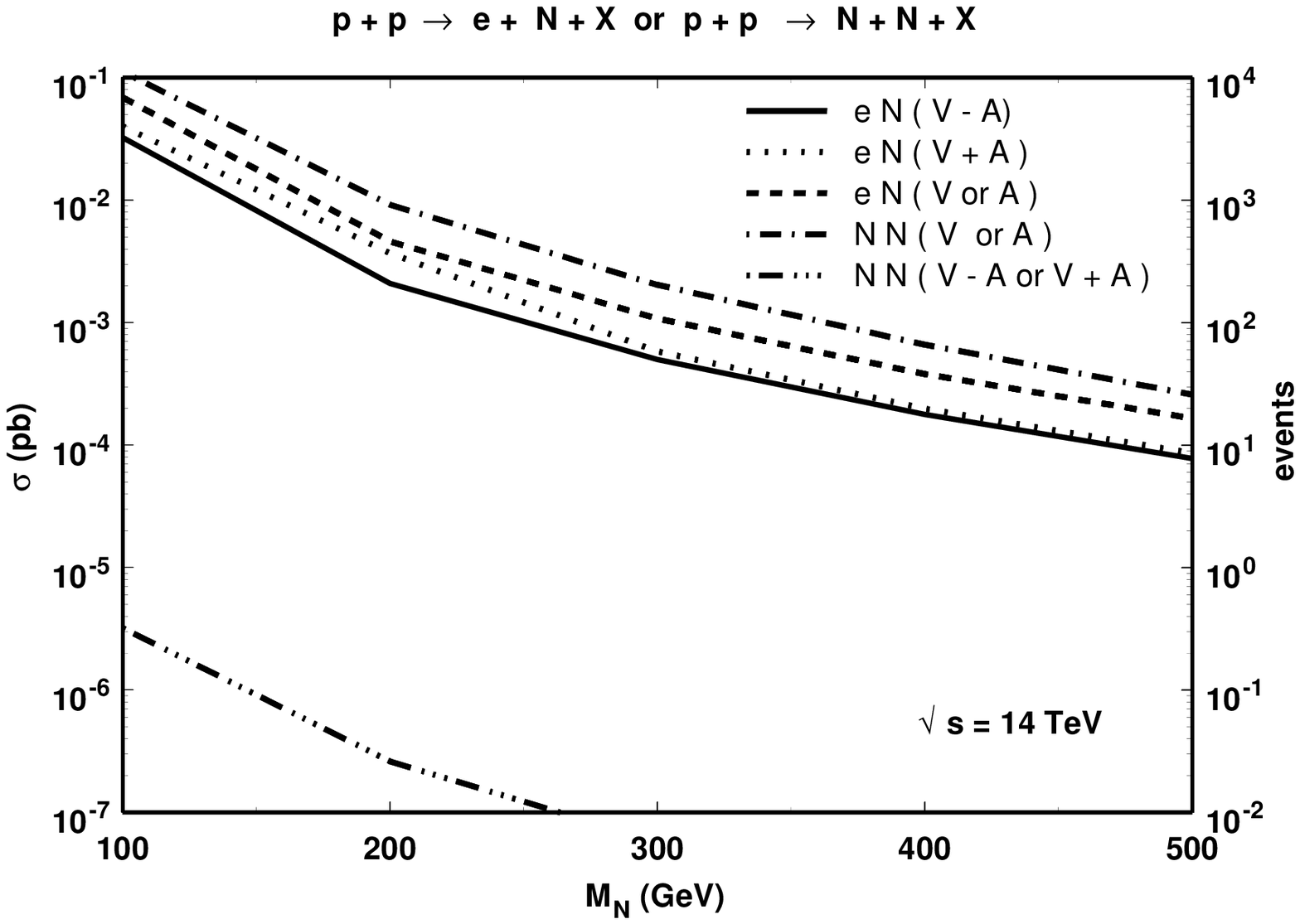}
\caption{Total cross section and number of events for $p + p \rightarrow e^- + N + X$ and  $p + p \rightarrow N + N + X$ for VSM, VDM, FMFM at $\sqrt s = 14$ TeV for ${\cal L} = 100$ fb$^{-1}$,
where $N$ stands for the new heavy Majorana neutrino.}
\end{figure} 
\par
The next important point is the nature of the Majorana couplings. Let us analyze first the real production of single Majorana neutrinos. For the process $p + p \longrightarrow e^- + N + X$ the primary electron rapidity could indicate the $V \mp A$ nature of the couplings, as shown in Figure 2. From these figures we can see clearly that there is no significant model dependence and the main differences come from the total cross-section.

\begin{figure} 
\includegraphics[width=0.5\textwidth]{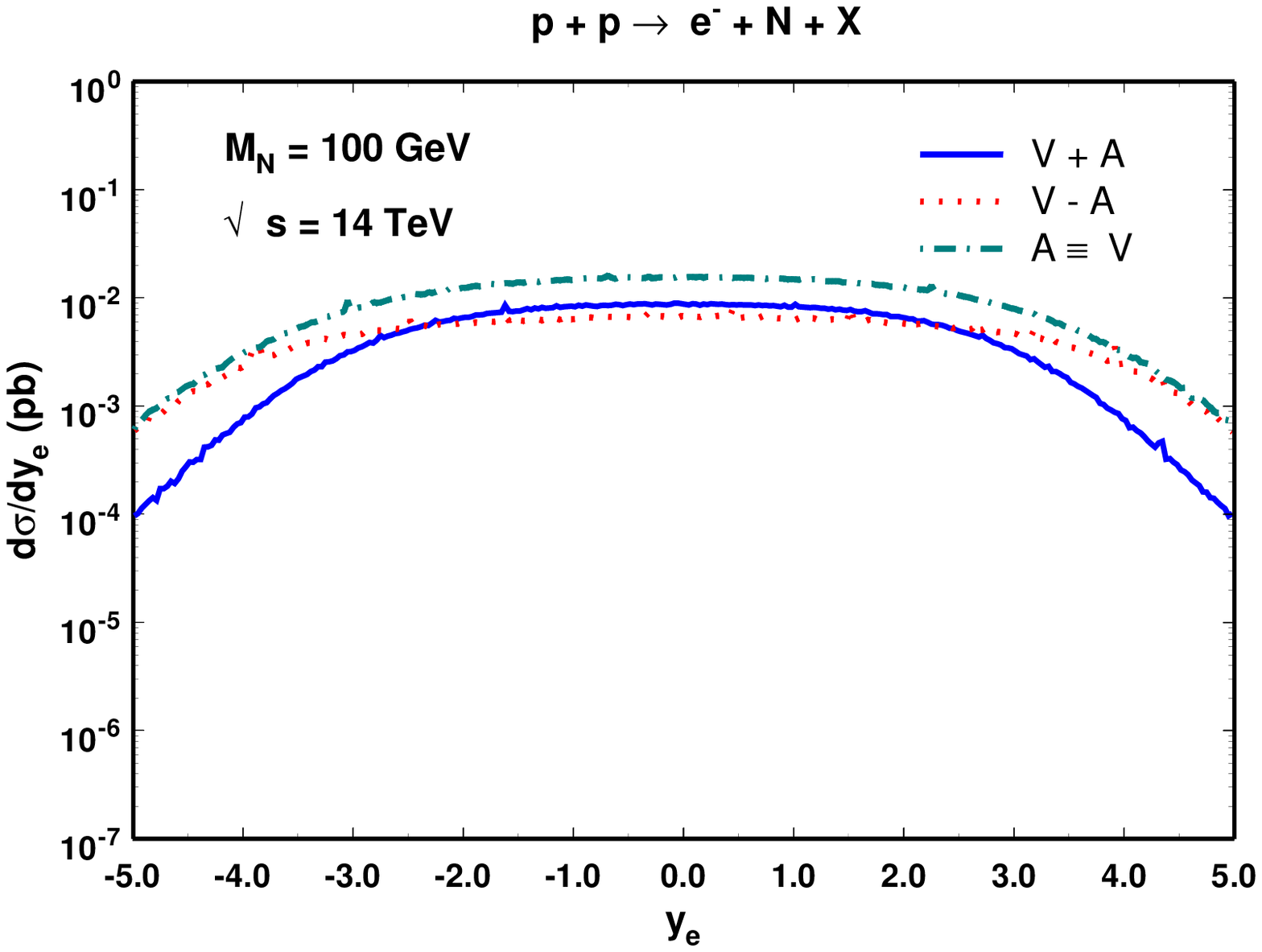}
\includegraphics[width=0.5\textwidth]{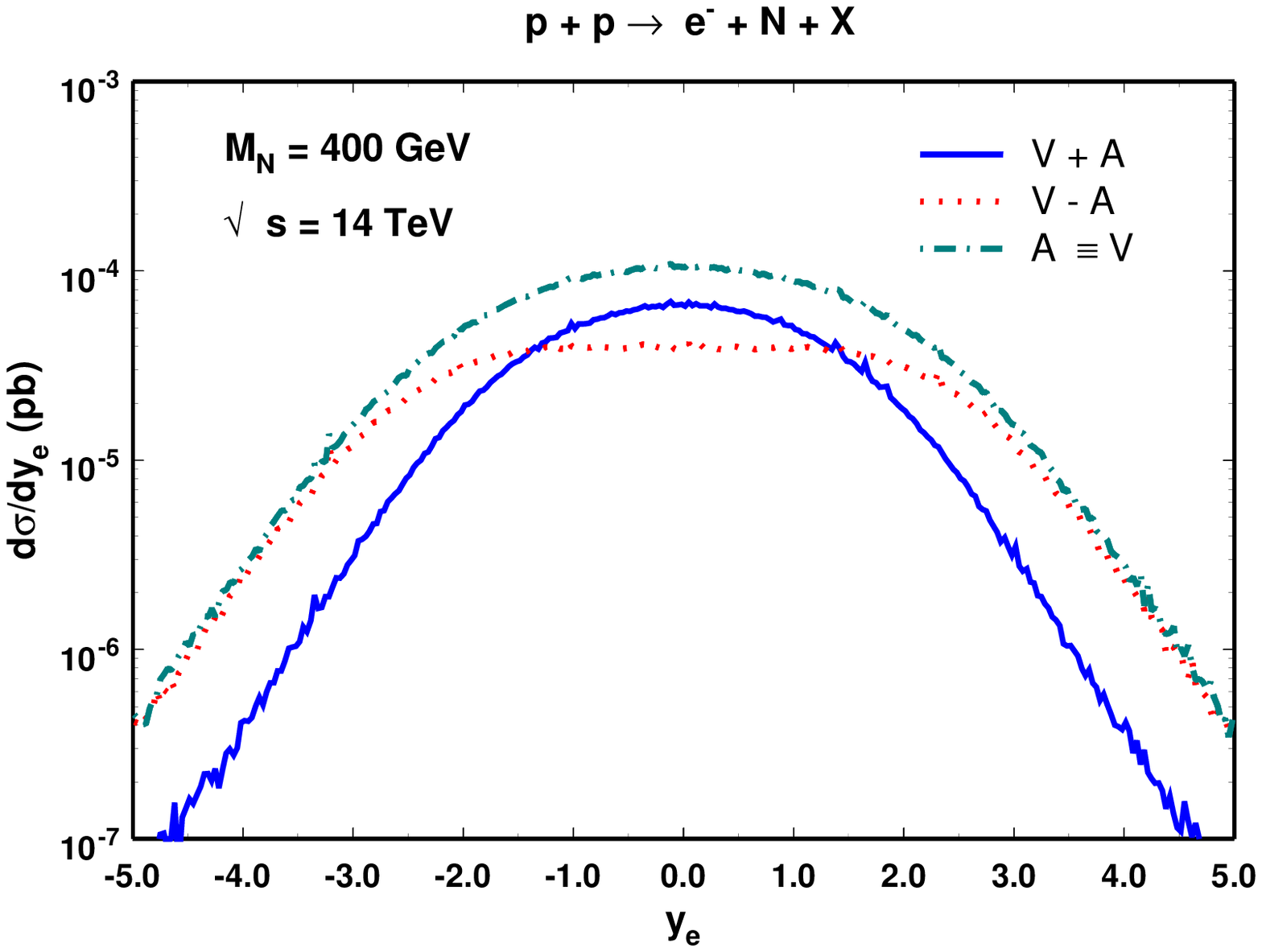}
\caption{Rapidity distributions for the final electron momentum in $p + p \longrightarrow e^- + N+ X$ for $M_N=100$ GeV (top) and $M_N=400$ GeV (bottom) for VSM, 
VDM, FMFM at $\sqrt s = 14$ TeV.}
\end{figure}

Since the heavy Majorana neutrino can decay as $N \longrightarrow e^{\mp} + W^{\pm}$, the most interesting, practical and clean signature is the process $p + p \longrightarrow  e^{\mp} + e^{\mp} + W^{\pm}+ X$, with two identical electrons or positrons in the final state. The rapidity of the final two electron momentum sum is given in Figure 3. In order to avoid double counting of the final identical particles we adopt to sum their momenta and calculate the rapidity of this sum.  These are more realistic distributions, for practical purposes, but again there is no strong model dependence in their shapes.

\begin{figure} 
\includegraphics[width=.5\textwidth]{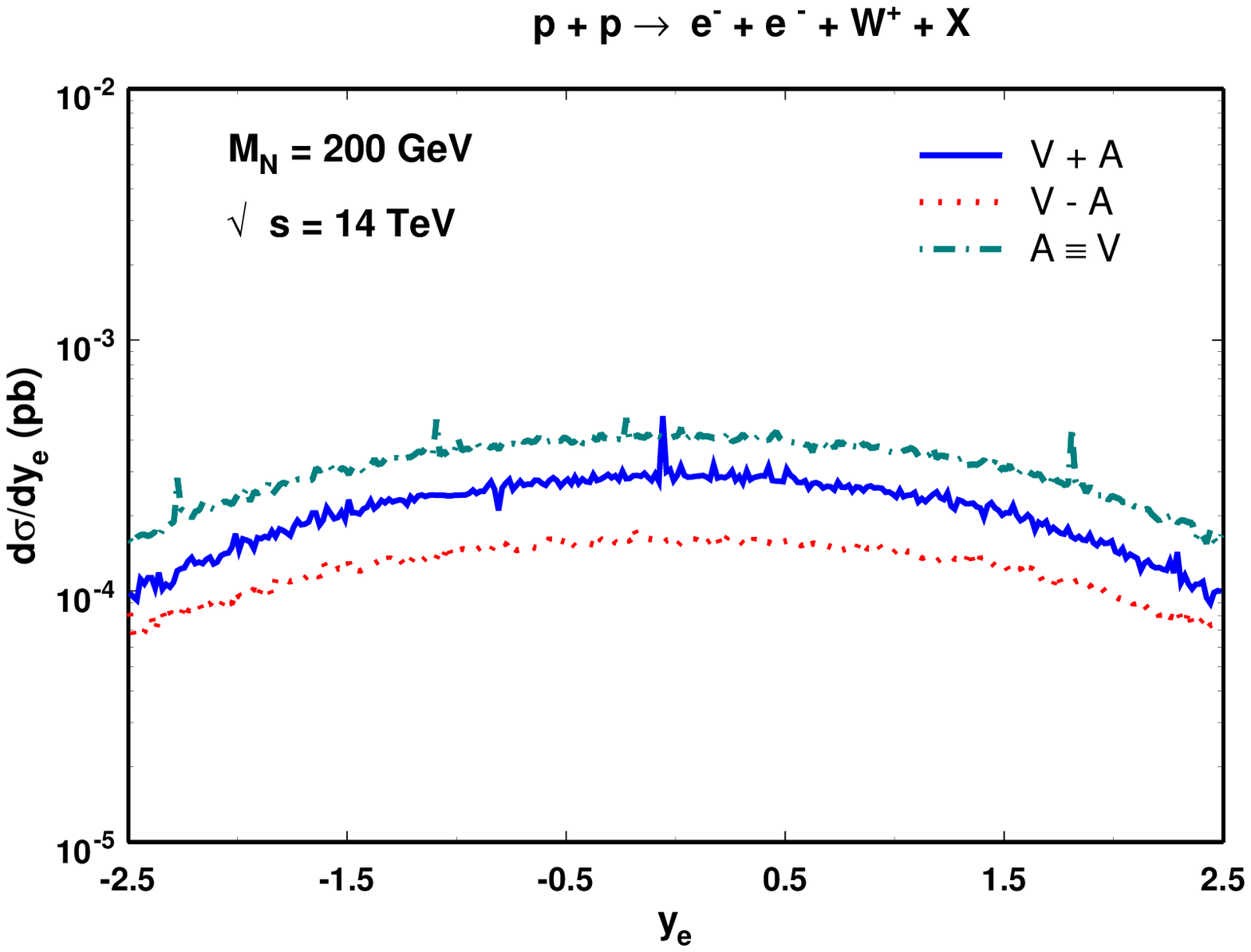}
\includegraphics[width=.5\textwidth]{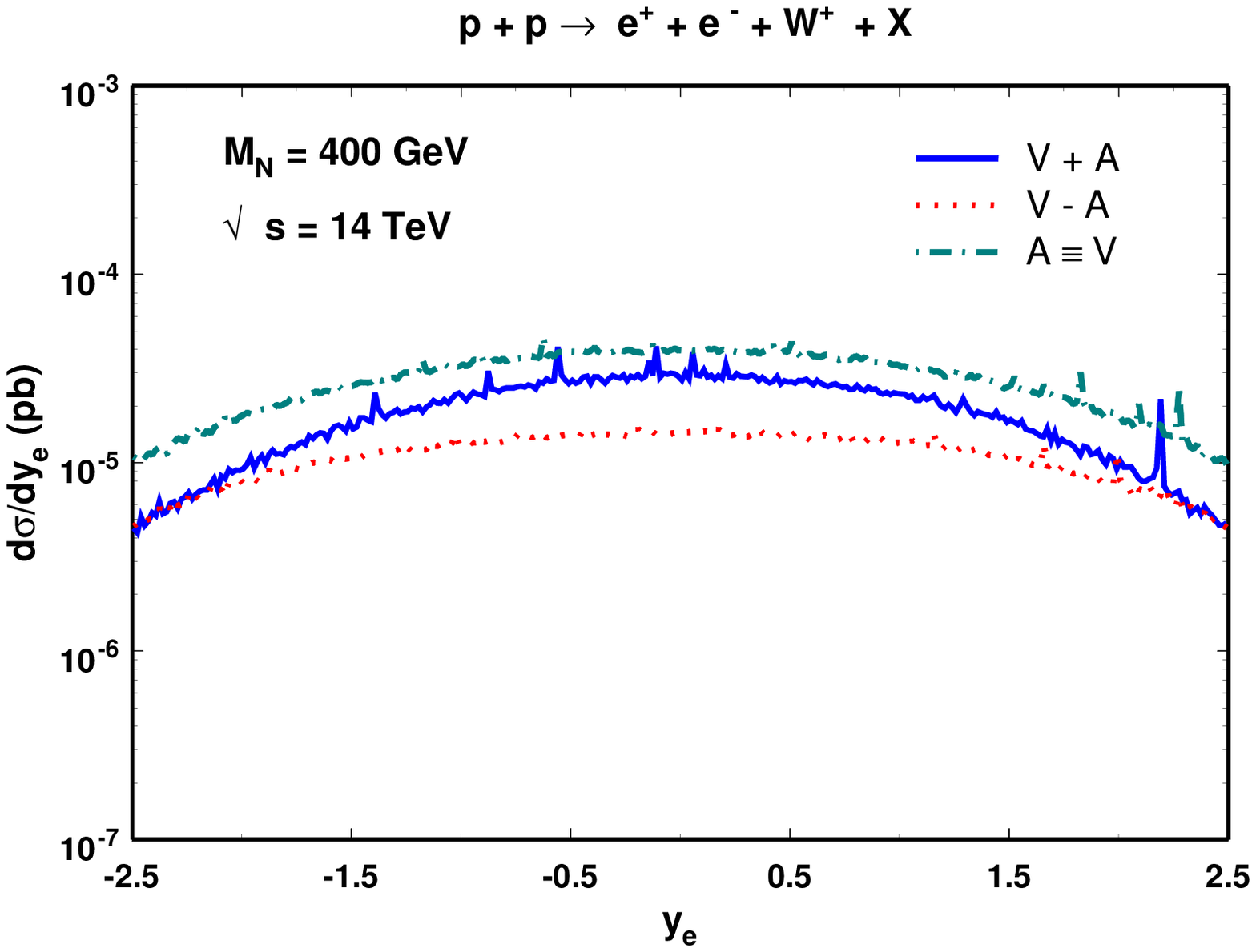}
\caption{Rapidity distributions for the sum of the final two electron momenta in $p + p \longrightarrow e^- + e^- + W^+ + X$ for $M_N=200$ GeV (top) and $M_N=400$ GeV (bottom) for VSM, VDM, FMFM at $\sqrt s = 14$ TeV.}
\end{figure}

In Table 2 we show the expected number of events for the
process $p + p \longrightarrow e^- + e^- + W^+ + X$ at $\sqrt s = 14$ TeV, for a luminosity of $100$ fb$^{-1}$, corresponding to one year of LHC operation at high luminosity regime. If we also consider the process $p + p \longrightarrow e^+ + e^+ + W^- + X$, the number of events in the table should be multiplied by two.

Although not so visually distinctive , model separation can be achieved through a statistical treatment using the rapidity of the lowest $p_T$ electron in the process $p + p \longrightarrow e^{\mp} + e^{\mp} + W^{\pm} + X$. The rapidity of highest $p_T$ electron showed to be less sensitive to characterize the difference between the models. The lowest $p_T$ electron distributions obtained for $M_N=100$ GeV and $M_N=200$ GeV for $\cal L$ = $300$ fb$^{-1}$ are showed in Figure 4 for the different models studied. One can see from Figure 4 (top) that it is not difficult to separate FMFM from the VSM and VDM, mainly due to the different cross sections, but the VSM and VDM separation is harder to do even for a $M_N=100$ GeV.  Figure 4 (bottom) shows that for $M_N= 200$ GeV it is already impossible to distinguish among the models, unless one uses a statistical treatment.

\begin{figure} 
\includegraphics[width=.5\textwidth]{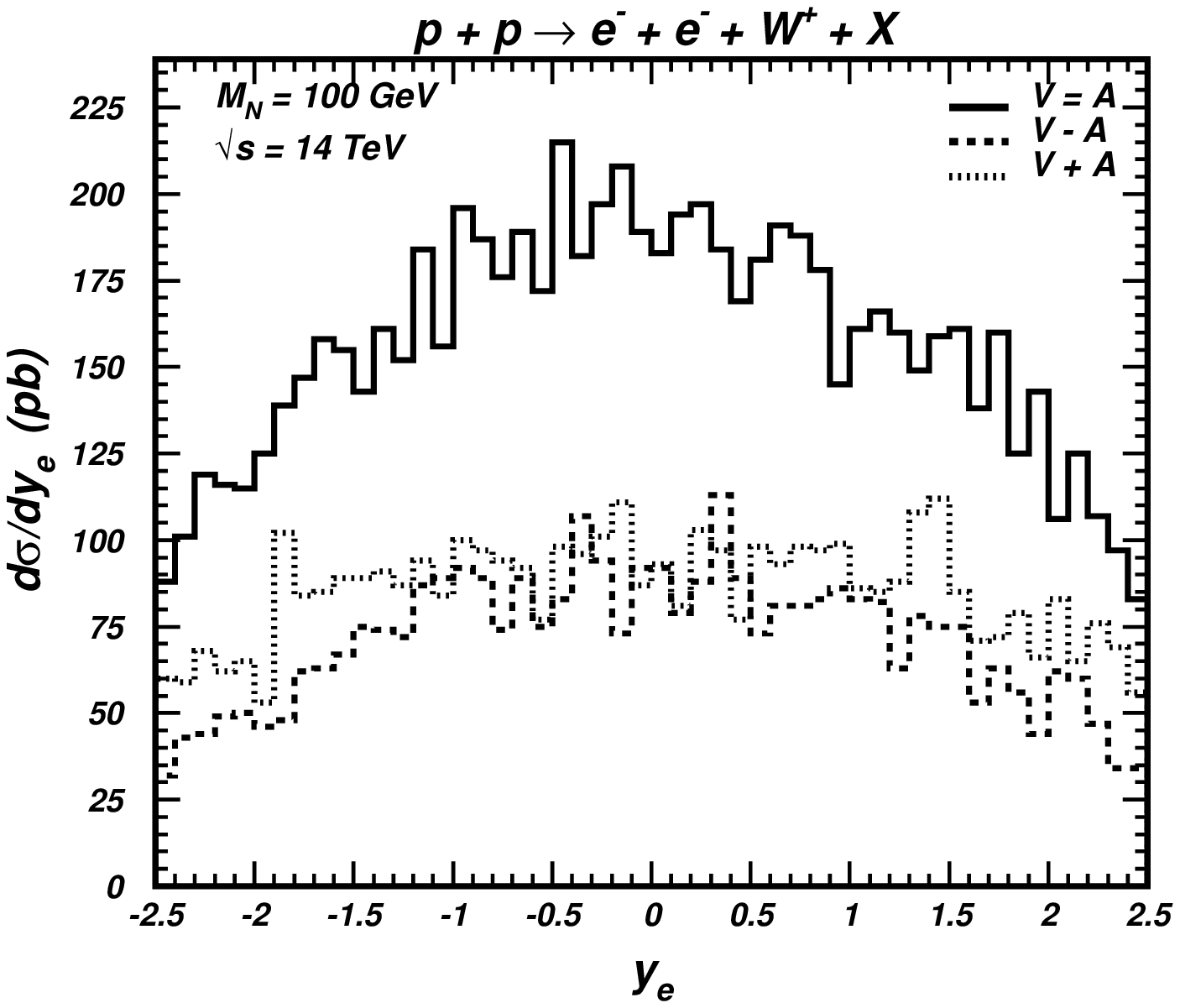}
\includegraphics[width=.5\textwidth]{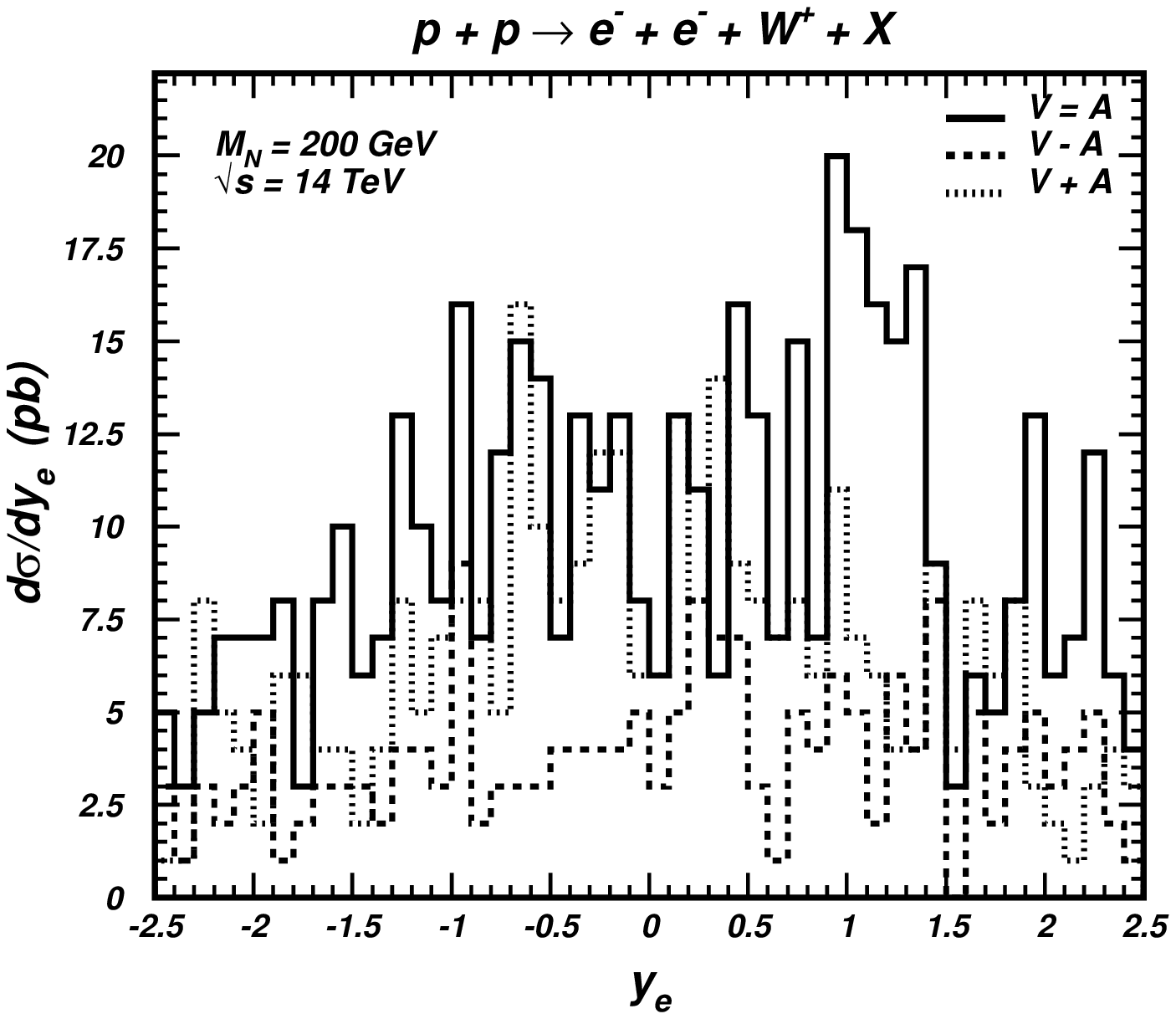}
\caption{Rapidity distributions for the final $e^-$ with lowest $p_T$ in $p + p \longrightarrow e^- + e^- + W^+ + X$ for $M_N=100$ GeV (top) and $M_N=200$ GeV (bottom) with ${\cal L} = 300$ fb$^{-1}$ for VSM, VDM, FMFM at $\sqrt s = 14$ TeV.}
\end{figure}

\begin{tablehere}\label{xuxa}
\begin{center}
\begin{footnotesize}
\begin{tabular}{|c|c|c|c|}
\hline \hline
$  $ &
\multicolumn{3}{|c|} {${\cal L}= 100fb^{-1}$}  \\ \hline
&   &   &   \\
$ M_N(GeV)$&   $VSM$ & $VDM$  & $FMFM$\\
&   &   &   \\
\hline
&   &   &   \\
$100$ & $1185$ &  $1426$ & 2614 \\
&   &   &    \\
\hline
&   &   &   \\
$200$  &   $63$  & $108$ & 161\\
&   &   &  \\ 
\hline
&   &   & \\
$300$ & $14$  & $22$ & 36 \\
&   &   &   \\ 
\hline
&   &   &   \\
$400$ &  $5$  &  $8$ & 13\\
&   &   &  \\ \hline
\hline
\end{tabular}
\end{footnotesize}
\caption{Number of events expected for the process $p + p \longrightarrow e^- + e^- + W^+ + X$ at $\sqrt s = 14$ TeV for VSM, VDM, and FMFM, for some Majorana neutrino masses and for  $\cal L$~=~$100$~fb$^{-1}$.} 
\end{center}
\end{tablehere}

\section{Model Distinction}

We performed a statistical treatment using an approximate $\chi^2$-function for Poisson distributions, adapting the Eq.(25) of reference~\cite{MBA} in order to compare the histograms and to find the Majorana mass limits allowing model separation for three different luminosities. We use the following $\chi^2$-function

\begin{equation}
\chi ^2 = \sum_{i=1}  2(n_i-k_i)+ (2n_i+1)\log\left(\frac{2n_i+1}{2k_i+1}\right),
\end{equation}
where $n_i$ and $k_i$ are the contents of the $i^{th}$ bin of each of the two histograms.

Our choice for Eq.(3) shows better results than the usual one \cite{PDG}, even in the cases when the number of events is small, especially when there are some bins with zero contents. For large number of events, Eq.(3) behaves asymptotically like the least square method. It converges faster, with much smaller number of events, and with less oscillations to the true parameter values when used for fitting data as shown in \cite{MBA}.  In order to be more realistic it was included an efficiency of 80\% for the electron identification and reconstruction.

\par
The events generated by CompHEP have been used to calculate the $\chi^2$-function differences among the histograms for different models and masses. The calculations have been done at the parton level and the final $W$ disintegration and hadronization have not been performed, since we are using only information from  the final electrons and they allow us a clean analysis. A clear signature is to require two identical electrons and hadron jets in the final state. Due to the low number of events and to analyze the histograms fluctuations, we have repeated the $\chi^2$-function calculations several times using $10$ different sets of data for each model, mass and luminosity and taken the mean values and their respective errors. This means that for each two model comparison, the $\chi^2$-function calculation has been done $100$ times in order to obtain the mean values and their respective errors.
Since we have two identical electrons in the final state, we have chosen the rapidity of the electron with the lowest $p_T$ to calculate the modified $\chi^2$. This variable distribution was shown to reinforce the differences among the models under study. We have not worried about the backgrounds, since this process is forbidden in the Standard Model and the background electrons coming from the decays of Standard Model particles have low energies.

The results for the separation among the models can be seen in Figure 5. The straight horizontal solid line represents the value above which models can be discriminated with a confidence level larger than $95\%$. This value has been calculated for $50$ degrees of freedom since we have used $50$ bins in the $<\chi^2>$-function calculations. In Figure 5 (top) we can see that for the VDM-FMFM combination, the models can be distinguished for $M_N$ above $250$ GeV, with data corresponding to one year of LHC operation at high luminosity. For a luminosity of $300$ fb$^{-1}$, VDM and FMFM can be separated if the heavy neutrino mass reaches $330$ GeV. For $500$ fb$^{-1}$, the separation between models can be achieved up to $M_N \simeq 400$ GeV. Similar mass bounds have been obtained for the separation between VSM and FMFM in Figure 5 (middle). On the other hand, for the VSM-VDM separation, heavy Majorana mass limits were found not so high, as shown in Figure 5 (bottom). This difference between the models reflects the different couplings ($V-A$, $V+A$, pure $V$ or pure $A$). The distinct total cross sections increase the differences in the rapidity distributions. For FMFM, no significant distinction was found between pure $V$ and pure $A$ couplings, as said before. In our calculations, FMFM was considered as pure $V$ couplings.

In order to better quantify  the above results, $p$-values are also calculated.
The $p$-value, in this paper, is the probability of a hypothesis given by a histogram $H_A$ be incompatible with another hypothesis given by histogram $H_B$. If the two histograms are strongly incompatible then $p \longrightarrow 0$, in the other hand if they are strongly compatible  $p \longrightarrow 1$.  The $p$-values are calculated according to 
\begin{equation}
p=\int_{<\chi^2>}^{\infty} f(z,n_d) dz
\end{equation}
where $f(z,n_d)$ is the $\chi^2$ probability distribution function and $n_d$ is the number of degrees of freedom, in this study $n_d=50$. $<\chi^2>$ are the values obtained when comparing the histograms and shown in Figures 5. The Tables III-V show the p-values obtained for different masses and luminosities. If we consider a model separation with 95\% C.L. then $p <0.05$. If the data at LHC show a compatibility with one of the three models described here, then it will be possible to distinguish the other two according to the Tables III-V. 
 
\par
\begin{figurehere} 
\includegraphics[width=.45\textwidth]{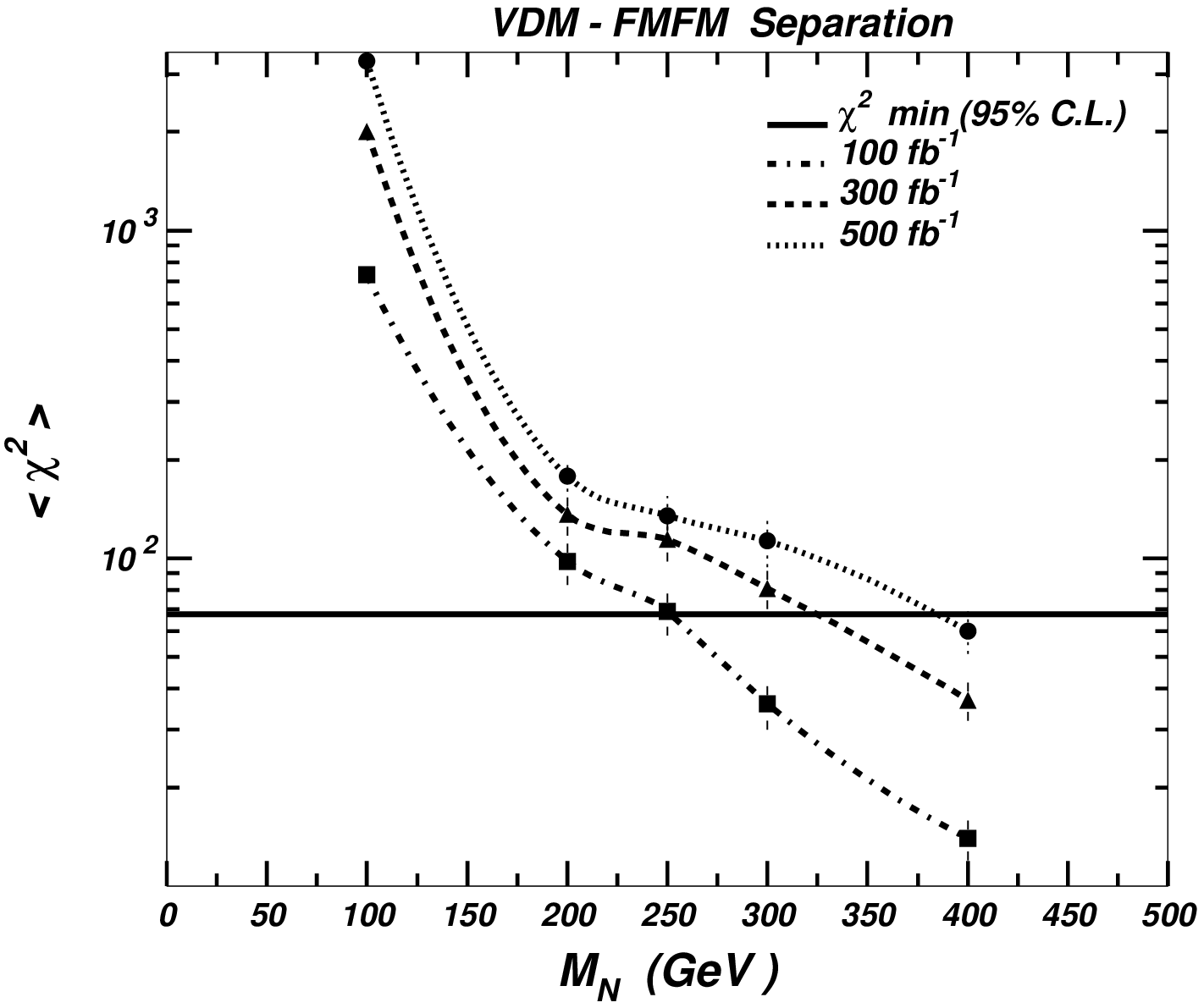}
\includegraphics[width=.45\textwidth]{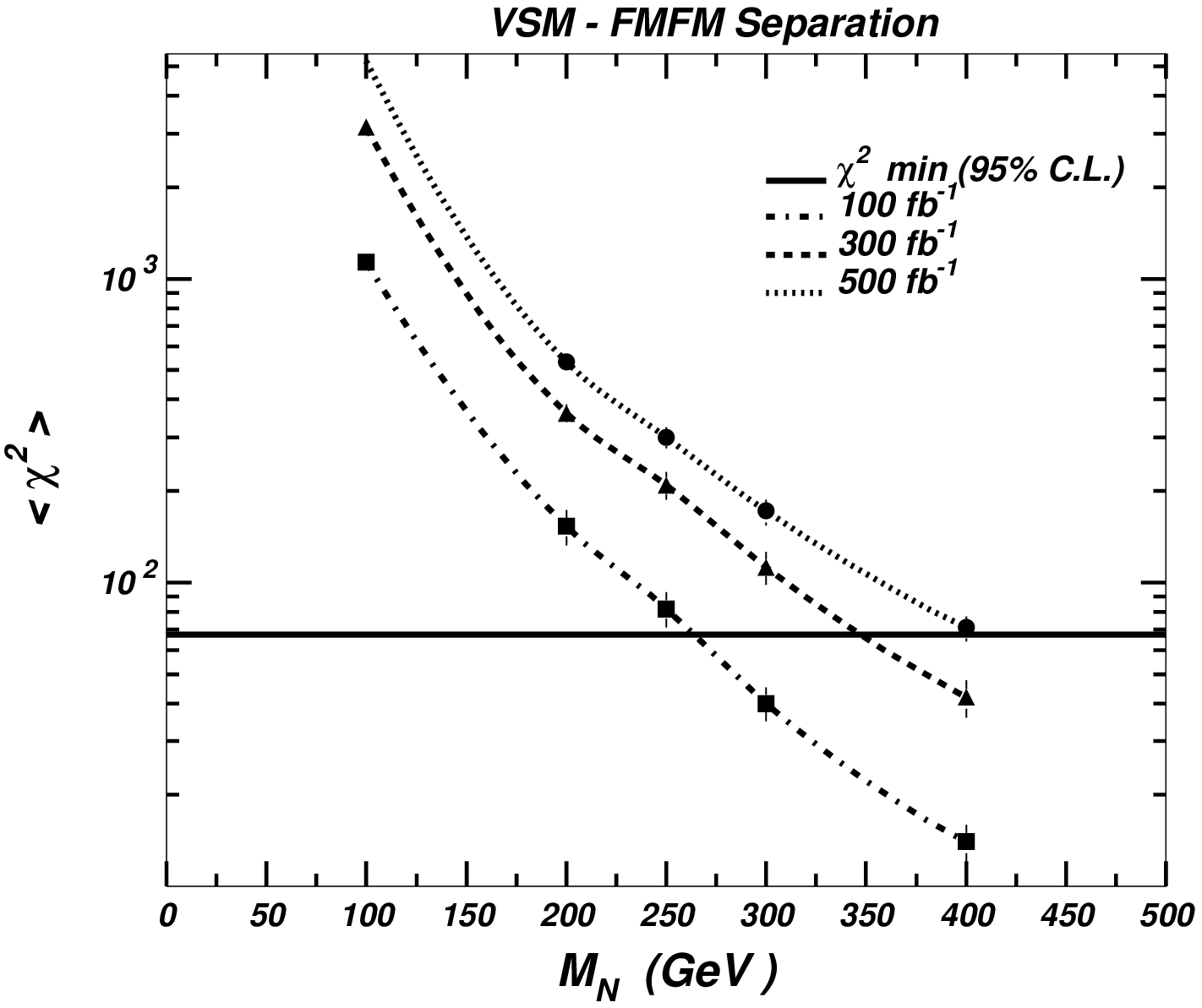} \\ 
\includegraphics[width=.45\textwidth]{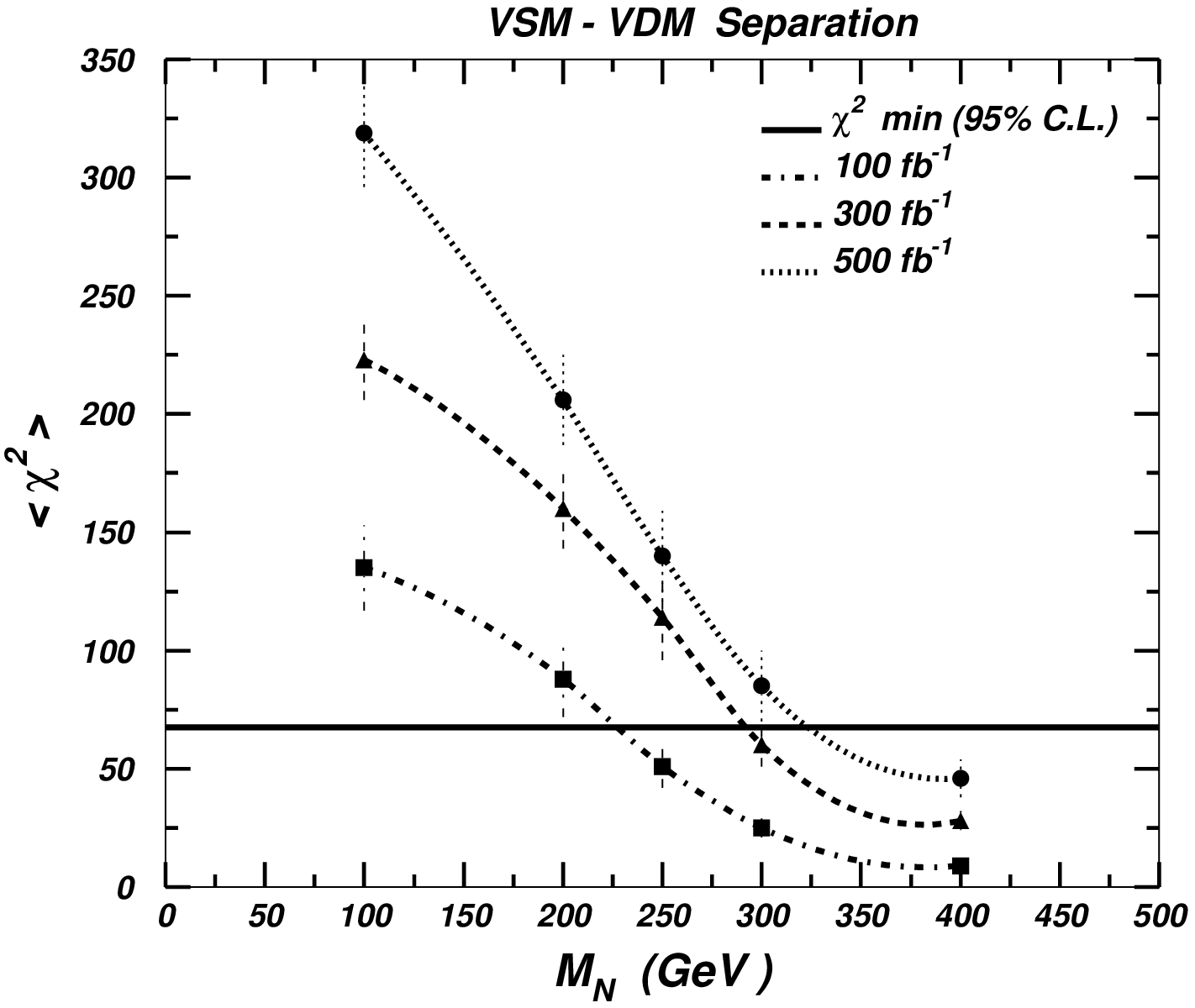}
\caption{Separation between the models VDM-FMFM (top), VSM-FMFM (middle) and VSM-VDM (bottom),  the horizontal line of each figure corresponds to $95\%$ C.L. based on the $<\chi^2>$ with $50$ degrees of freedom. Note the logarithmic vertical scale for the top two figures.}
\end{figurehere}

\bigskip

\begin{tablehere}\label{xuxa4}
\begin{footnotesize}
\begin{tabular}{|c||c|c|c|}
\hline \hline
$M_N$ & ${\cal L}= 100fb^{-1}$ & ${\cal L}= 300fb^{-1}$ & ${\cal L}= 500fb^{-1}$ \\
$(GeV)$ &  &  &  \\
\hline
 \hline
100  &  $0.102651\cdot 10^{-120}$ &  $0.275233\cdot 10^{-387}$  &   $0.42576 \cdot 10^{-660}$ \\
\hline
200  & $0.588073 \cdot 10^{-4} $ &   $0.691069 \cdot 10^{-9}$ &  $0.29823 \cdot 10^{-15}$ \\
\hline
250  &  $0.386516 \cdot 10^{-1}$ &   $0.664087\cdot 10^{-6}$ &  $0.95806 \cdot 10^{-9}$ \\
\hline
 300  &  $0.931739$ &  $0.360976\cdot 10^{-2}$  &  $0.891443 \cdot 10^{-6}$ \\
\hline
400  &  $0.999999$ &  $0.913923$ &  $0.157242$ \\
\hline
\hline
\end{tabular}
\end{footnotesize}
\caption{$p$-values for different masses and luminosities when comparing VDM and FMFM} 
\end{tablehere}
\bigskip

\begin{tablehere}\label{xuxa3}
\begin{footnotesize}
\begin{tabular}{|c||c|c|c|}
\hline \hline
$M_N$ & ${\cal L}= 100fb^{-1}$ & ${\cal L}= 300fb^{-1}$ & ${\cal L}= 500fb^{-1}$ \\
$(GeV)$ &  &  &  \\
\hline
\hline
100  & $0.304084 \cdot 10^{-203}$ & $0.233734 \cdot 10^{-633}$ & $0.434935 \cdot 10^{-1073}$ \\
\hline
200  & $0.224681\cdot 10^{-11}$  & $0.108177 \cdot 10^{-47}$  &  $0.837551\cdot 10^{-81}$  \\
\hline
250  & $0.289768\cdot 10^{-2}$  & $0.247785 \cdot 10^{-20}$   &   $0.231413\cdot 10^{-36} $\\
\hline
 300  &  $0.843227$ &  $0.119454\cdot 10^{-5}$ &  $0.266346\cdot 10^{-14}$ \\
\hline
400  &  $0.999999$ &   $0.782155$  &   $0.270090 \cdot 10^{-1}$ \\
\hline
\hline
\end{tabular}
\end{footnotesize}
\caption{Same as Table III but for VSM and FMFM} 
\end{tablehere}
\bigskip

\begin{tablehere}\label{xuxa2}
\begin{footnotesize}
\begin{tabular}{|c||c|c|c|}
\hline \hline
$M_N$ & ${\cal L}= 100fb^{-1}$ & ${\cal L}= 300fb^{-1}$ & ${\cal L}= 500fb^{-1}$ \\
$(GeV)$ &  &  &  \\
 \hline
\hline
$100$  & $0.958060 \cdot 10^{-9}$ & $0.105196 \cdot 10^{-22}$ &  $0.747901 \cdot 10^{-40}$  \\
\hline
$200$  & $0.728142\cdot 10^{-3}$ & $0.194823 \cdot 10^{-12}$   &  $0.7882270 \cdot 10^{-20}$ \\
\hline
$250$  & $0.434093$   & $0.664087 \cdot 10^{-6}$ &  $0.184826 \cdot 10^{-9}$ \\
\hline
 $300$  & $0.998807$ & $0.157242$  &  $0.1471710 \cdot 10^{-2}$ \\
\hline
$400$  &  $0.999999$  &  $0.994980$  &  $0.6345808$ \\
\hline
\hline
\end{tabular}
\end{footnotesize}
\caption{Same as Table III but for VSM and VDM} 
\end{tablehere}

\section{Conclusions}

\par
This work suggests that the process $p + p \longrightarrow  e^{\mp} + e^{\mp} + W^{\pm} + X$ is an adequate, relatively simple and clear channel to look for a heavy Majorana neutrino and also to study its origin at hadronic colliders, since it violates leptonic number conservation rule and has, {\it a priori}, no background from the Standard Model. For an annual luminosity of $100$ fb$^{-1}$ this channel allows an investigation of the existence of a Majorana neutrino up to a mass of $500$ GeV. Furthermore, we showed that, once experimentally found, the theoretical origin of the Majorana neutrino can be determined using the rapidity distribution of the final electron with the lowest $p_T$, provided there are enough data. Although the differences among the rapidity distributions for the several models studied are not so great, we showed that a model separation could be achieved using a special statistical treatment, until a mass of the order of $300-350$ GeV for $300$~fb$^{-1}$ of data, at the parton level. The use of the lowest $p_T$ electron rapidity distribution enhances the difference among the models, total cross sections and thus allows model discrimination until higher Majorana masses. A treatment, considering the hadronic jets resulting from the $W$ disintegration, could lead to model separation for higher heavy neutrino Majorana masses. The LHC experimental groups can do a more realistic treatment doing full detector simulation. Due to the few number of parameters involved in the models studied here and the clear non-SM channel, we believe this kind of experimental analysis can be done soon with data from the first "runs".
\vskip 1cm
{\it Acknowledgments:} This work was partially supported by the
following Brazilian agencies: CNPq and FAPERJ. The work of M. A. B. do Vale has been partially supported by the PRONEX project research grants from FAPEMIG and CNPq.

}
\end{document}